\input{psfig.sty}
\documentstyle[tighten,aps]{revtex}

\begin{document}


\title{\bf {Extended Kinetic Models with Waiting-Time Distributions:\\
         Exact Results}}
\author{Anatoly B. Kolomeisky\footnote{Now at Department of Chemistry, Rice University, Houston, Texas 77005-1892}  and Michael E. Fisher \\
Institute for Physical Science and Technology,\\
University of Maryland, College Park, \ MD 20742 USA}
\date{\today}

\maketitle
\pagenumbering{arabic}

\begin{abstract} 
\hspace{1em} Inspired by the need for effective stochastic  models  to   describe   the complex behavior of biological  motor proteins that move on linear tracks, exact results are derived  for the velocity and dispersion of  simple linear  sequential models (or one-dimensional random walks)  with general {\it waiting-time distributions}. The concept of ``mechanicity'' is introduced in order to conveniently quantify departures from simple ``chemical,'' kinetic rate processes,  and its significance  is briefly  indicated.  The results are extended to  more elaborate  models that have finite side-branches and include death processes (to represent the detachment of a motor from the track).

\end{abstract}



\section {\hspace{0.5em}Introduction and Summary}

Motor proteins such as kinesins, dyneins, myosins, DNA and RNA polymerases, are important   for the  biological functioning of  cells. Consuming  energy obtained  from the hydrolysis of ATP or related compounds, and moving along rigid linear tracks (microtubules,  actin filaments, DNA molecules, etc.),  they play significant roles in cell division, cellular transport, muscle contraction and genetic transcription$^{1,2}$.  Such molecular motors can move with  velocities, $V$, up to 1000 nm/s$^{3-6}$  and may sustain an external load, $F$, of 5-8 pN for kinesins$^{6}$  and up to 30-40 pN for DNA and RNA polymerases$^{7,8}$.  Understanding   the detailed  mechanism of the functioning of motor proteins is a major  challenge  of modern biology.

In recent years,  significant  advances have been made in experimental techniques for  studying   motor proteins: one can now  observe and  investigate  accurately  the mechanical  properties of single   molecules  over wide parameter ranges$^{3-8}$. However, our theoretical understanding of how these proteins work  is still incomplete. Theoretical modeling of molecular  motors    has followed    two main  directions. One class of models  is based on ``thermal ratchets''$^{9-12}$  in which the motor protein is viewed as a  Brownian particle that diffuses in two or more periodic but  asymmetric  potentials between which it switches stochastically. Another, more traditional ``chemical''  approach is based on multistate   kinetic descriptions of the motion with various rate processes determining the transitions between the states$^{13-23}$.  In this paper we consider various  extensions of the chemical kinetic schemes for which we derive exact, closed form results in terms of the underlying transition rate parameters.

The simplest  periodic sequential  kinetic model assumes that a motor protein molecule moves along a  periodic molecular track (a microtubule in the case of a  kinesin) and binds at sites at $x=ld$ $(l=0, \pm 1, \pm 2, \cdots)$. It is postulated that there are $N$ discrete states, $j=0,1,\cdots,N-1$,  between consecutive binding sites and the motor protein in  state $j_{l}$ (``at'' site $l$) can move forward to state $(j+1)_{l}$ (normally pulling a load)  at a  rate $u_{j}$ and can slip backward, to state $(j-1)_{l}$  at a  rate $w_{j}$$^{18,19}$.  This basic $N$-state model is, clearly, precisely isomorphic to a discrete but, in general,  biased random walk on a periodic one-dimensional lattice. Such random walks are of broad interest for a variety of applications$^{24-29}$, in particular, for studying diffusion in random environments$^{28-30}$ which may, for example, be approached by allowing the period $N$ to become infinite$^{28}$. Indeed, some time ago Derrida \cite{derrida} presented a mathematical approach that provides formally exact and explicit formulas for the asymptotic drift velocity
\begin{equation}
V_{0}=V_{0}(\{u_{j},w_{j}\})=\lim_{t \rightarrow \infty} \frac{d}{dt}\langle x(t) \rangle,
\end{equation}
and  for the  dispersion (or  effective diffusion constant)
\begin{equation}
D_{0}=D_{0}(\{u_{j},w_{j}\})=\case{1}{2}\lim_{t \rightarrow \infty}\frac{d}{dt}\left [\langle x^{2}(t) \rangle - \langle x(t) \rangle^{2} \right ],
\end{equation}
where  $x(t)$ is  the spatial position of the motor protein molecule or, equally, of the random walker  along the linear track at time $t$.  Derrida's results have recently been exploited in developing theory for molecular motors$^{17-20,23}$. A valuable feature of the exact closed expressions is that it is quick and straightforward to explore the precise effects of a wide range of parameter values, revealing a variety of different characteristic types of behavior$^{18-20}$, in contrast to employing  approximate numerical schemes  or Monte Carlo simulations, or to studying   over-simplified  models .

The ``purpose'' of a motor protein in a biological setting is to move various biochemical entities,  such as vesicles, etc., that exert load forces, $F$, which  act on the protein molecule. The questions  of what  force can be  exerted by a  motor  protein  within the basic model and how the transition rates should be changed by the load, $F$, have been discussed   critically$^{18,19}$. The resulting  calculations for these models  show qualitative and good semi-quantitative agreement with the  experimentally observed behavior of normal  two-headed kinesins$^{3-8,18,19,23}$.

In order to describe  motor proteins more realistically, however,  and to understand  the behavior of proteins other than two-headed kinesins$^{7,8,31,32}$, extensions of the basic $N$-state  periodic sequential  kinetic model were recently  introduced$^{20}$. These  extended  models   take into consideration the complexity of real  biochemical cycles (multiple kinetic paths  and branched states) and allow for the  irreversible stochastic  detachment of molecular motors from the linear track which is always observed. Exact expressions for velocities and dispersions have been derived for these extended chemical kinetic models$^{20}$ in terms of the basic forward and backward rate constants, $u_{j}$ and $w_{j}$, the corresponding rate constants for the side branches, and the death rates from any of the $N$ intermediate states.

Now, the basic concept underlying  chemical kinetic models is the idea that the motion  of the motor protein   is essentially ``chemical,''  i.e., the molecule undergoes a transition or ``jump''  from one  chemical state to a  nearest neighbor state (in the kinetic diagram) at a  given rate having ``forgotten'' how it arrived at the state.  The time intervals  between such jumps are thus  distributed exponentially according to Poisson statistics.  In other words,  the time intervals are described by   exponential {\it waiting-time  distribution} functions in which  the coefficient in  the exponent represents  the overall rate of  transition from  the  state in question. 

A conceptually possible  alternative picture of the motion is  that it is  ``purely  mechanical'' so that  the motor works  like  a clock,  jumping  from state to state within  narrowly distributed time intervals.  Real   molecular motors work  with a surprisingly  high mechanical  efficiency  (which may be estimated as 80-90$\%$ for kinesins). This,  in turn,  suggests that some of  the steps in  the dynamical sequence  of the motion  might reasonably be described as rather more  mechanical than purely random in nature. In that case, the waiting-time distribution functions for the time intervals marking transitions  between different internal states might be modeled  more effectively or more economically$^{23}$  as  {\it non}-exponential.  To that end we present here a generalization of  the basic  periodic chemical kinetic models, and of  their extensions,  that  explicitly  considers  {\it general}  waiting-time  distributions. By using the results of Montroll and coworkers$^{24-27}$  we are, again, able to obtain explicit, exact results for $V$ and $D$ for the various models. We also  explore briefly  the quantitative  significance  of the  ``mechanical'' factors that could represent departures from purely chemical kinetic processes  in the motion of molecular motors. 

To proceed, consider first  the  extension of the simple sequential kinetic model (with period $N$)  shown  schematically in   \mbox{Fig. 1}. To characterize the dynamics of the motor protein (or random walkers) we introduce waiting-time  distribution functions $\psi_{j}^{\pm}(t)$ and  $\psi_{j}^{0}(t)$,  where $\psi_{j}^{+}(t) dt$ is the probability of jumping one  step forward from state $j$ in the  time interval  $t$ to $t+dt$ after arriving in state $j$, while  $\psi_{j}^{-}(t) dt$ is the corresponding  probability of jumping one step backward from state $j$,  and $\psi_{j}^{0}(t) dt$ is the   probability of attempting to move in the same time interval  but failing to do so and hence restarting at the same site  after the attempt. The allowance for $\psi_{j}^{0}(t)$ is potentially useful in modeling motor proteins since it can describe, say, the ``futile'' hydrolysis of an ATP molecule without achieving a forward step. 

A more complicated model allows for  the possibility of irreversible detachments  or ``deaths''. Deaths are described by  a distribution $\psi_{j}^{\delta}(t)$ such that  $\psi_{j}^{\delta}(t) dt$   is the probability of leaving  from state $j$ in the time interval  $t$ to $t+dt$ but not appearing in any other state$^{20}$. Alternatively, the detachment or death process may carry the motor, or walker, to a reservoir (or graveyard) from which no returns are allowed.

 Finally,  another possible extension of the simple kinetic sequential scheme is to incorporate  finite side-branch processes$^{20}$: see \mbox{Fig. 2}. If state $(j,k)_{l}$ labels  a  state $k$ on a branch emanating from the primary   state $j_{l}$,  the dynamics  can be   described by additional distributions $\psi_{j,k}^{\beta}(t)$ where   $\psi_{j,k}^{\beta}(t) dt$  is the probability of jumping one  step further out  from the branch state $(j,k)_{l}$ in the  time interval  $t$ to $t+dt$ ($k=0,1,\cdots,L-1$), while  $\psi_{j,k}^{\gamma}(t) dt$ is the corresponding  probability of jumping one  step back in towards the primary state $j_{l} \equiv (j,0)_{l}$ from the branch state $(j,k)_{l}$ ($k=1,2,\cdots,L$). Note that without loss of generality we may  assume  that all side branches are  of the same  length $L$: see \mbox{Ref. 20}.

The waiting-time distribution functions respect the basic $N$-state periodicity so that 
\begin{equation}
\psi_{j,k}^{\zeta}(t)=\psi_{j \pm N,k}^{\zeta}(t),
\end{equation}
where  $\zeta=+,-,0,\delta,\beta,$ or $\gamma$.  Furthermore, for any state $(j,k)$   normalization requires
\begin{equation}
\sum_{\zeta} \int_{0}^{\infty}\psi_{j,k}^{\zeta}(t)dt=1.
\end{equation}

For all these extensions of the original chemical kinetic model we have derived  explicit general  expressions for the drift velocities and for the dispersions by generalizing  Derrida's method$^{28}$. When appropriate   exponential waiting-time distribution functions are substituted,  we recover    Derrida's original  formulas$^{28}$  for the simple linear model  and, likewise,  our  previous  results for the   extended  kinetic models$^{20}$. For convenience, we present all  our concrete results,  namely, the expressions for the mean  velocities and for the dispersions,  in this section. The detailed, and unavoidably somewhat involved,  calculations and derivations  are described   in \mbox{Sections III-V}.

To quantify the importance of ``purely  mechanical'' factors in the motion of motor proteins the concept of {\it mechanicity} is introduced and discussed in \mbox{Section  II}. The  mechanicity  varies from 0 for a ``chemical'' or  Poisson process (with exponential waiting-time distribution functions) to 1 for a purely  mechanical process (with  clock-work or delta-function waiting-time distributions).  It provides  a convenient  quantitative measure of deviations from a simple chemical picture of the dynamics of  molecular motors which may well  prove useful in applications$^{23}$. 

To simplify the presentation of our  results it is convenient to introduce the  {\it overall waiting-time distribution},   $\psi_{j,k}(t)$, which is merely  the sum of all the  distinct  distributions associated  with  the state $(j,k)$. (Recall that $k=0$ corresponds to primary states, $j_{l}\equiv (j,0)_{l}$, on the linear sequence).  We will need the  Laplace transform of the overall distribution defined  by
\begin{eqnarray}
\widetilde{\psi}_{j,k}(s)&=& \int_{0}^{\infty} e^{-st}\psi_{j,k}(t) dt \nonumber\\
&=& 1-s \mu_{1;j,k} + \case{1}{2} s^{2} \mu_{2;j,k} - \cdots,
\end{eqnarray}
where $\mu_{n;j,k}$ ($n=0,1,\cdots$) is the $n$-th moment of $\psi_{j,k}(t)$ and we have used the normalization condition (4) to conclude $\mu_{0;j,k} \equiv 1$. The transforms, $\widetilde{\psi}_{j,k}^{\zeta}(s)$, of the partial waiting-time distributions and their moments $\mu_{n;j,k}^{\zeta}$, are defined similarly. Of course, for the  specific  models that we consider  some of the  $\widetilde{\psi}_{j,k}^{\zeta}$  will be absent (and, so, can be set to zero).

The analysis reveals that a crucial role is played by the  associated  {\it waiting-time rate  distributions},  $\varphi_{j,k}^{\zeta}(t)$, which are  defined via their   Laplace transforms
\begin{equation}
\widetilde{\varphi}_{j,k}^{\zeta}(s) \equiv  s \widetilde{\psi}_{j,k}^{\zeta}(s)/[1-\widetilde{\psi}_{j,k}(s)],
\end{equation}
where $\zeta$, as before,  can be $+,-,0,\delta,\beta$ or $\gamma$. Note that it is the transform $\widetilde{\psi}_{j,k}$ of the total or  overall waiting-time distribution that appears here in the denominator.   

It then transpires, as seen below and proved in \mbox{Sections III-V}, that in terms of the $\widetilde{\varphi}_{j,k}^{\zeta}(s)$ one can readily define {\it effective kinetic transition rates}:   $\ u_{j}$ and $w_{j}$, for the forward and backward rates along the primary linear sequence; $\ \delta_{j}$ for the death processes from states $j \equiv (j,0)$; and side-branch rates, $\beta_{j,k}$ and $\gamma_{j,k}$, outward and inward from the states $(j,k)$.  In the main, especially for the velocity $V[\{\psi_{j,k}^{\zeta}(t)\}]$, these effective kinetic rates play the same role in our results as do the simple (Poissonian) rates, denoted by the same symbols$^{18-20}$, in the analysis of the original kinetic models with exponential waiting-time distributions.  Furthermore, for the velocity and dispersion (in the absence of death processes), only low order moments of the  $\varphi_{j,k}^{\zeta}(t)$  enter explicitly. These, in turn can be expressed in terms of low order waiting-time moments, $\mu_{n;j,k}$ and   $\mu_{n;j,k}^{\zeta}$. To be explicit, dropping the state labels $(j,k)$, we can write
\begin{equation}                                                                          \widetilde{\varphi}^{\zeta}(s)=v_{0}^{\zeta} -sv_{1}^{\zeta} + \case{1}{2} s^{2} v_{2}^{\zeta} - \cdots ,        
\end{equation}
where the rate moments $v_{n}^{\zeta}$ are expressed in terms of the $\mu_{n}$ and $\mu_{n}^{\zeta}$ in \mbox{Table I}. We will assume that all the moments displayed in the Table are finite.

As found previously in analyzing  the various simple  kinetic models, our  expressions for $V$ and $D$  in the  extended  models with waiting-time  distributions depend on certain  linear sequential products of rate ratios. Following$^{20}$, we thus  define  two  types of product for the   sequential model with waiting times, namely, 
\begin{equation}
\Pi_{j}^{k} \equiv \prod_{i=j}^{k}\frac{w_{i}}{u_{i}} \quad \mbox{and} \quad \Pi_{j}^{\dagger k} \equiv \prod_{i=j}^{k}\frac{w_{i+1}}{u_{i}}=\frac{w_{k+1}}{w_{j}}\Pi_{j}^{k} .
\end{equation}
The model with branches requires one more type, namely,  the branch products 
\begin{equation}
\Pi_{j}^{\beta,k} \equiv \prod_{i=1}^{k}\frac{\beta_{j,i-1}}{\gamma_{j,i}}.
\end{equation}
For the model with deaths or  irreversible detachments we introduce  modified analogs of (8), namely,  $\widetilde{\Pi}_{j}^{k}$ and $\widetilde{\Pi}_{j}^{\dagger k}$, that are obtained simply  by substituting   $u_{j}$ and $w_{j}$  by the  ``renormalized'' values
\begin{equation}
\tilde{u}_{j}=u_{j}\varepsilon_{j+1}/\varepsilon_{j} \quad \mbox{ and } \quad \widetilde{w}_{j}=u_{j}\varepsilon_{j-1}/\varepsilon_{j},
\end{equation}
where the periodic  renormalization coefficients, $\varepsilon_{j} \equiv\varepsilon_{j \pm N}$ ($j=0,1,\cdots,N-1$), are conveniently   normalized  by  the condition 
\begin{equation}
\varepsilon_{0} \equiv 1.
\end{equation}
The remaining  coefficients  $\varepsilon_{j}$  then form the  components of the right eigenvector, $\mbox{\boldmath{$\varepsilon$}} =[\varepsilon_{j}]$, corresponding to  the smallest  eigenvalue, $\lambda=\lambda(\{u_{j},w_{j}\})$,  of the  $N$$\times$$N$  transition rate matrix ${\bf M}[\{u_{j},w_{j};\delta_{j}\}]$   which is defined by the nonzero elements
\begin{equation}
M_{j,j-1}=-w_{j},  \quad  M_{j,j}=u_{j}+w_{j}+\delta_{j}, \quad  M_{j,j+1}=-u_{j},
\end{equation}
and, because of periodicity,
\begin{equation}
M_{0,-1} \equiv M_{0,N-1}=-w_{0} \quad  \mbox{and} \quad M_{N-1,N} \equiv M_{N-1,0}=-u_{N-1}.
\end{equation}

With these preliminaries established we can now present our explicit results in fairly compact form.

\subsection{\hspace{0.5em} Results for the  Simple Sequential Model with Waiting Times }

The formal expression for the velocity in this case is precisely  the same as derived by Derrida$^{28}$  for the sequential chemical  kinetic model, that is
\begin{equation}
V=d(1-\Pi_{1}^{N})/R_{N}, 
\end{equation}
where, using the definitions given above, 
\begin{equation}
R_{N}=\sum_{j=0}^{N-1} r_{j}, \quad \quad \quad r_{j}=u_{j}^{-1}\left[ 1+\sum_{k=1}^{N-1} \Pi_{j+1}^{j+k}\right ],
\end{equation}
and $d$ is a distance between neighboring binding sites   on the linear track. However,  the effective transition rates $u_{j}$ and $w_{j}$ are now given by
\begin{equation}
u_{j}, \hspace{0.5em} w_{j} =  \widetilde{\varphi}_{j}^{\pm}(s=0)=v_{0;j}^{\pm}=\frac{\int_{0}^{\infty}\psi_{j}^{\pm}(t) dt}{\int_{0}^{\infty}t \psi_{j}(t) dt},
\end{equation}
where we have used (7) and \mbox{Table I} and may recall that $ \hspace{0.5em} \psi_{j}=\psi_{j}^{+}+\psi_{j}^{-}+\psi_{j}^{0}$. 

The dispersion takes a new, more complex form which can be written
\begin{equation}
D=D_{0}+D_{1},
\end{equation}
where the  first term is  given by
\begin{equation}
D_{0}=(d/N)\{[VS_{N}+d U_{N}]/(R_{N})^{2}-\case{1}{2}(N+2)V\},
\end{equation}
\begin{equation}
S_{N}=\sum_{j=0}^{N-1} s_{j} \sum_{k=0}^{N-1}(k+1) r_{k+j+1}, \quad \quad U_{N}=\sum_{j=0}^{N-1} u_{j} r_{j} s_{j},
\end{equation}
in which new coefficients  $s_{j}$ are  determined by
\begin{equation}
s_{j}=u_{j}^{-1}\left[ 1+\sum_{k=1}^{N-1} \Pi_{j-1}^{\dagger j-k} \right].
\end{equation}
The second contribution  to the  dispersion   is found  to be
\begin{equation}
D_{1}=(d/N) [NV/(R_{N})^{2}] \sum_{j=0}^{N-1} s_{j}(g_{j}^{+}r_{j}-g_{j+1}^{-}r_{j+1}),
\end{equation}
where new, ``nonexponential parameters'',  $g_{j}^{\pm}$, are defined by
\begin{equation}
g_{j}^{\pm} = \left. \frac{d \widetilde{\varphi}_{j}^{\pm}}{d s} \right| _{s=0}=-\int_{0}^{\infty} t \varphi_{j}^{\pm}(t) dt=-v_{1;j}^{\pm}.
\end{equation}
For the relation of these parameters to the waiting-time moments, see \mbox{Table I}. When  the  waiting times are  exponentially distributed with, in particular, $\psi_{j}^{+}(t)$ , $\psi_{j}^{-}(t)$, and  $\psi_{j}^{0}(t)$ each  proportional to $\exp(-c_{j}t)$,  the parameters  $g_{j}^{\pm}$ vanish identically.  When this occurs for all $j$, the dispersion  $D$ is given by $D_{0}$ alone and, as mentioned,   Derrida's  original  formula  is recovered$^{18,19,28}$. We remark that the $g_{j}^{\pm}$ are typically negative but may, in fact, also be positive: their character is discussed in more concrete terms in \mbox{Section II}.

\subsection{\hspace{0.5em}Sequential Model with Branches and   Waiting-Times}

\renewcommand{\thefootnote}{\arabic{footnote}}

For the models  with  branches of finite length (see \mbox{Fig. 2})  the velocity is given by
\begin{equation}
V_{\beta}(\beta,\gamma)=d(1-\Pi_{1}^{N})/R_{N}^{\beta}, 
\end{equation}
which   is identical  to the expression (14) for the unbranched  models  except for the  modified functions
\begin{equation}
R_{N}^{\beta}=\sum_{j=0}^{N-1} r_{j}^{\beta}, \quad \quad r_{j}^{\beta}=r_{j}[1+ \sum_{k=1}^{L} \Pi_{j}^{\beta,k}],
\end{equation}
in which  the effective transition rates in and out of the  branch states are defined  in parallel to (16) by
\begin{equation}
\beta_{j,k} =  \widetilde{\varphi}_{j,k}^{\beta}(s=0), \quad \quad \gamma_{j,k} =  \widetilde{\varphi}_{j,k}^{\gamma}(s=0).
\end{equation}
It is appropriate to recall however, that the rates $\beta_{j,0}$, $u_{j}$ and $w_{j}$  involve $\psi_{j,0}(t)$ which now entails the {\it three} distributions $\psi_{j}^{+}$, $\psi_{j}^{-}$ and  $\psi_{j,0}^{\beta}$.

The dispersion can be expressed as
\begin{equation}
D_{\beta}=D_{0,\beta}+D_{1,\beta}+D_{2,\beta},
\end{equation}
where  the first two  terms are  very similar to $D_{0}$ and $D_{1}$ for the unbranched  models:  explicitly we find
\begin{equation}
D_{0,\beta}=(d/N)\{[V_{\beta}S_{N}^{\beta}+d U_{N}^{\beta}]/(R_{N}^{\beta})^{2}-\case{1}{2}(N+2)V_{\beta}\},
\end{equation}
\begin{equation}
D_{1,\beta}=(d/N) [NV_{\beta}/(R_{N}^{\beta})^{2}] \sum_{j=0}^{N-1} s_{j}^{\beta}(g_{j}^{+}r_{j}-g_{j+1}^{-}r_{j+1}),
\end{equation}
where the modified functions, analogous to $S_{N}$ and $U_{N}$ in (19), are 
\begin{equation}
S_{N}^{\beta}=\sum_{j=0}^{N-1} s_{j}^{\beta} \sum_{k=0}^{N-1}(k+1) r_{k+j+1}^{\beta}, \quad \quad U_{N}^{\beta}=\sum_{j=0}^{N-1} u_{j} r_{j} s_{j}^{\beta},
\end{equation}
in which, in parallel to (20), we require
\begin{equation}
s_{j}^{\beta}=u_{j}^{-1} \left[ 1+\sum_{l=1}^{L} \Pi_{j}^{\beta,l}+\sum_{k=1}^{N-1}\left(1+\sum_{l=1}^{L} \Pi_{j-k}^{\beta,l}\right)\Pi_{j-k}^{\dagger j-1} \right].
\end{equation}
The third term in (26), which is a contribution   due to the presence of branches, is given by
\begin{equation}
D_{2,\beta}=V_{\beta}^{2} \sum_{j=0}^{N-1} \sum_{k=1}^{L} W_{j,k}^{\beta},
\end{equation}
where, for the  coefficients, we have
\begin{eqnarray}
W_{j,k}^{\beta} &=&\frac{r_{j}}{R_{N}^{\beta} \gamma_{j,k}} \left[ \sum_{l=k}^{L} \Pi_{j}^{\beta,l} - g_{j,k-1}^{\beta}\Pi_{j}^{\beta,k-1} + g_{j,k}^{\gamma}\Pi_{j}^{\beta,k} \right. \nonumber \\
& &  \left.  + \sum_{i=1}^{k-1}\frac{\beta_{j,i}}{\beta_{j,0}}\Pi_{j}^{\beta,i} \left(  \sum_{l=k-i}^{L} \Pi_{j}^{\beta,l} - g_{j,k-i-1}^{\beta}\Pi_{j}^{\beta,k-i-1} + g_{j,k-i}^{\gamma}\Pi_{j}^{\beta,k-i} \right) \right].
\end{eqnarray}
The nonexponential parameters,  $g_{j,k}^{\beta}$ and $g_{j,k}^{\gamma}$, are defined in precise  analogy to (22) by
\begin{equation}
g_{j,k}^{\beta} = (d \widetilde{\varphi}_{j,k}^{\beta}/d s) |_{s=0}=-v_{1;j,k}^{\beta}, \quad \quad g_{j,k}^{\gamma} = (d \widetilde{\varphi}_{j,k}^{\gamma}/d s) | _{s=0}=-v_{1;j,k}^{\gamma},
\end{equation}
where, again, the relation to the waiting-time moments follows from \mbox{Table I}. Similarly, if $\psi_{j,k}^{\zeta}(t)=Q_{j,k}^{\zeta} \exp(-c_{j,k}t)$ holds for all $(j,k)$, the results of \mbox{Ref. 20} are once more obtained. \mbox{Section II} discusses expressions for the $g_{j,k}^{\zeta}$ in terms of associated ``mechanicities'' $M_{j,k}^{\zeta}$.

\subsection{\hspace{0.5em}Sequential Model with Deaths  and   Waiting-Times}

The mean velocity is now  given  by 
\begin{equation}
V_{\delta}=d(1-\widetilde{\Pi}_{1}^{N})/{R}_{N}^{\delta}, 
\end{equation}
where $\widetilde{\Pi}_{1}^{N}$   and the $\widetilde{r}_{j}$ (appearing below)  are defined, using  $\widetilde{u}_{j}$ and $\widetilde{w}_{j}$ [see (10)], in exact  analogy to (8) and (15). However, in contrast to  (16),  the effective transition rates $u_{j}$ and  $w_{j}$ and the death rate  $\delta_{j}$, which enter the expression for $R_{N}^{\delta}$, require new definitions. Specifically,  we find 
\begin{equation}
u_{j}, \hspace{0.5em} w_{j} \equiv  \widetilde{\varphi}_{j}^{\pm}(s=-\lambda)=\frac{\lambda \int_{0}^{\infty} e^{+\lambda t} \psi_{j}^{\pm}(t) dt}{\int_{0}^{\infty} (e^{+\lambda t}-1) \psi_{j}(t) dt},
\end{equation}
and $\delta_{j}=\widetilde{\varphi}_{j}^{\delta}(s$=$-\lambda)$ which can, likewise, be expressed in the integral form exhibited in (35), where it may be recalled that $\psi_{j}=\psi_{j}^{+}+ \psi_{j}^{-}+\psi_{j}^{0}+\psi_{j}^{\delta}$. In these expressions   $\lambda$ is, as stated above,   the smallest eigenvalue of the  transition rate matrix ${\bf M}[\{u_{j},w_{j};\delta_{j}\}]$ [see (12) and (13)]. The function ${R}_{N}^{\delta}$ is then  given by
\begin{equation}
R_{N}^{\delta}=\sum_{j=0}^{N-1} r_{j}^{\delta}, \quad \quad  r_{j}^{\delta}=[1+(1-\alpha_{j})g_{j}^{\delta}] \widetilde{r}_{j},
\end{equation}
where the modified death and non-exponential parameters $g_{j}^{\delta}$ and $g_{j}^{\pm}$ are embodied in
\begin{equation}
\alpha_{j} \equiv [g_{j}^{+}(\varepsilon_{j+1}-\varepsilon_{j})+g_{j}^{-}(\varepsilon_{j-1}-\varepsilon_{j})]/(\varepsilon_{j} g_{j}^{\delta}),
\end{equation}
while the parameters themselves  are now given by
\begin{equation}
g_{j}^{\delta} \equiv (d \widetilde{\varphi}_{j}^{\delta}/d s) | _{s=-\lambda}, \quad \quad g_{j}^{\pm} \equiv (d \widetilde{\varphi}_{j}^{\pm}/d s) | _{s=-\lambda},
\end{equation}
where the integral expressions corresponding to that in (35) are somewhat more elaborate but may be found straitforwardly from (6) and will mirror the form of $v_{1}^{\zeta}$ in \mbox{Table I}. Recall from (10)-(13) that the coefficients $\varepsilon_{j}$ form the right eigenvector of ${\bf M}$ for $\lambda$, the smallest  eigenvalue.

Finally, the  dispersion can be written as
\begin{equation}
D_{\delta}=D_{0,\delta}+D_{1,\delta}+D_{2,\delta},
\end{equation}
where the  first term is given by
\begin{equation}
D_{0,\delta}=(d/N)\{[V_{\delta}S_{N}^{\delta}+d U_{N}^{\delta}]/(R_{N}^{\delta})^{2}-\case{1}{2}(N+2)V_{\delta}\},
\end{equation}
in which  $S_{N}^{\delta}$ and  $U_{N}^{\delta}$  are defined  in precise analogy to (19), that is,
\begin{equation}
S_{N}^{\delta}=\sum_{j=0}^{N-1} s_{j}^{\delta} \sum_{k=0}^{N-1}(k+1) r_{k+j+1}^{\delta}, \quad \quad U_{N}^{\delta}=\sum_{j=0}^{N-1} \widetilde{u}_{j} \widetilde{r}_{j} s_{j}^{\delta},
\end{equation}
$\widetilde{r}_{j}$ being defined just  after (34), while the analog of (20) is
\begin{equation}
s_{j}^{\delta}=\widetilde{u}_{j}^{-1}\left[ 1+ (1-\alpha_{j})g_{j}^{\delta}+\sum_{k=1}^{N-1}[1+(1-\alpha_{j-k})g_{j-k}^{\delta}] \Pi_{j-1}^{\dagger j-k} \right].
\end{equation}
The second term in (39) is then  given by
\begin{equation}
D_{1,\delta}=(d/N) [NV_{\delta}/(R_{N}^{\delta})^{2}] \sum_{j=0}^{N-1} s_{j}^{\delta}(\widetilde{g}_{j}^{+}\widetilde{r}_{j}-\widetilde{g}_{j+1}^{-}\widetilde{r}_{j+1}),
\end{equation}
where we have introduced  the renormalized modified nonexponential parameters
\begin{equation}
\tilde{g}_{j}^{\pm}=g_{j}^{\pm}\varepsilon_{j \pm 1}/\varepsilon_{j}.
\end{equation}
Finally, the last term in the  dispersion, which arises solely because of the possibility of death or  detachment,  is given by
\begin{equation}
D_{2,\delta}=\case{1}{2} (V_{\delta}^{2}/R_{N}^{\delta}) \sum_{j=0}^{N-1}(1-\beta_{j}) h_{j}^{\delta} \widetilde{r}_{j},
\end{equation}
where the  new functions 
\begin{equation}
\beta_{j} = [h_{j}^{+}(\varepsilon_{j+1}-\varepsilon_{j})+h_{j}^{-}(\varepsilon_{j-1}-\varepsilon_{j})]/(\varepsilon_{j} h_{j}^{\delta}),
\end{equation}
incorporate the second-order nonexponential parameters 
\begin{equation}
h_{j}^{\pm} = (d^{2} \widetilde{\varphi}_{j}^{\pm}/d s^{2}) | _{s=-\lambda}=\int_{0}^{\infty}t^{2} e^{\lambda t} \varphi_{j}^{\pm}(t)  dt,
\end{equation}
\begin{equation}
h_{j}^{\delta} = (d^{2} \widetilde{\varphi}_{j}^{\delta}/d s^{2}) | _{s=-\lambda} = \int_{0}^{\infty}t^{2} e^{\lambda t} \varphi_{j}^{\delta}(t)  dt.
\end{equation}
It is interesting that these higher moments of the  waiting-time rates,  $\varphi_{j}^{\pm}(t)$ and  $\varphi_{j}^{\delta}(t)$,  arise only when death processes come into play. They can be written as integrals of the corresponding $\psi_{j}^{\zeta}(t)$, as in (35), but will then exhibit the structure of $v_{2}^{\zeta}$ in \mbox{Table I} and entail, in particular, the modified third moments $\int_{0}^{\infty} t^{3} e^{\lambda t} \psi_{j}^{\pm}  dt$, etc. Note that when the probability of death or detachment vanishes, one has  $\lambda=0$ and  the coefficients  $\alpha_{j}, \beta_{j}, h_{j}^{\pm}, h_{j}^{\delta}$ and $g_{j}^{\delta}$ all vanish; then we recapture the results stated above for simple sequential model with waiting times.

\section {\hspace{0.5em}Degrees of Mechanicity}

Our analysis of the velocity and dispersion for the  stochastic models with waiting-time distributions has revealed that in terms of the effective rates $u_{j}$, $w_{j}$, etc. [as defined in (16), (25), and (35)] deviations from Poisson processes with nonexponetial waiting times do {\it not} change the formal expressions for the velocities: compare, e.g., (14), (23) and (34) with the corresponding  results presented in \mbox{Ref. 20}. On the other hand, the expressions for the dispersions change dramatically and, in particular, involve new ``nonexponetial parameters'' like $g_{j}^{\pm}$ [see (22), (33), (39), etc.] that evidently provide a measure of some sort  for the departures from simple ``chemical,'' kinetic processes. In order to gain a more concrete and intuitive picture of what these departures involve, we introduce a quantitative concept which we call the {\it mechanicity}, $M_{j}^{\zeta}$, of the specific transition process described by the waiting-time distribution $\psi_{j}^{\zeta}(t)$ (where, as previously, $\zeta=+,-,0,\cdots$).

The basic idea is to discriminate, in an explicit way, a standard exponential distribution from distributions that depart from it in varying degrees, and from a sharp, ``clockwork'' distribution of zero width. Now if $\overline{t}$ is the mean waiting time for a particular process, the mean square deviation $\overline{(\Delta t)^{2}}=\overline{t^{2}}-\overline{t}^{\ 2}$ provides a natural measure of the width of the distribution. Then the dimensionless ratio $\Theta = \overline{(\Delta t)^{2}}/\overline{t}^{\ 2}$ represents a scale-free index of the relative width. In the purely mechanical or clockwork limit $\Delta t= t- \overline{t}$ must vanish identically and so $\Theta=0$; conversely, if the distribution $\psi(t)$ yielding $\overline{t}$ and $\overline{(\Delta t)^{2}}$ is a simple exponential, one has $\Theta \equiv 1$. Thus the mechanicity parameter $M \equiv 1-\Theta$ vanishes for a Poisson or ``chemical'' process but attains the value unity for a purely mechanical process.

More illustratively$^{18}$, suppose the waiting-time distribution has the familiar general form
\begin{equation}
\psi(t)=Q t^{\nu -1} e^{-\sigma t} \quad \quad (\nu >0).
\end{equation}
An elementary calculation then yields
\begin{equation}
M \equiv 1- \Theta = 1- \nu^{-1}.
\end{equation}
By construction, the chemical limit is described by $\nu=1$; conversely, the mechanical limit is realized when $\nu \rightarrow \infty$. This example  also shows that $M$ may be {\it negative} (and, then, indefinitely large). Evidently, this arises when $\psi (t)$ is sharply peaked at the origin exhibiting a power law behavior $\sim 1/t^{\chi}$ with $\chi=|M|/(1+|M|)$ (for $M<0$).

In applying these considerations to a general waiting-time distribution, $\psi_{j}^{\zeta}(t)$, with moments
\begin{equation}
\langle t_{\zeta}^{n} \rangle_{j} \equiv \mu_{n;j}^{\zeta}= \int_{0}^{\infty} t^{n} \psi_{j}^{\zeta} (t) dt,
\end{equation}
a little care is needed since the zeroth moment, $\mu_{0;j}^{\zeta} \equiv \langle t_{\zeta}^{0} \rangle_{j}$, represents the total probability that the transition of type $\zeta$ occurs which, in general, is less than unity. The appropriate definition of the mechanicity for the process $\zeta$ from a state $j$ is thus
\begin{equation}
M_{j}^{\zeta} \equiv 1- \Theta_{j}^{\zeta} = 2-\langle t_{\zeta}^{0} \rangle_{j} \langle t_{\zeta}^{2} \rangle_{j}/\langle t_{\zeta} \rangle_{j}^{2}.
\end{equation}

Now in applying the various expressions displayed in \mbox{Section I} for the dispersions, $D$, it would be convenient if the nonexponential parameters, $g_{j}^{\pm}$ (or $g_{j}^{\beta}$ and $g_{j}^{\gamma}$), could be expressed directly in terms of the associated mechanicities, $M_{j}^{\pm}$ (and/or $M_{j}^{\beta}$ and $M_{j}^{\gamma}$) and the effective rates
\begin{equation}
u_{j} = \langle t_{+}^{0} \rangle_{j}/ \tau_{j} \quad \mbox{ and } w_{j} = \langle t_{-}^{0} \rangle_{j}/ \tau_{j}
\end{equation}
(and/or $\beta_{j,k}$ and $\gamma_{j,k}$), where if, for simplicity, we suppose only forward and backward processes act from state $j$, the mean dwell time is just
\begin{equation}
\tau_{j}=\langle t_{+} \rangle_{j}+\langle t_{-} \rangle_{j}=(u_{j}+w_{j})^{-1}.
\end{equation}
However, because the $g_{j,k}^{\zeta}$ are defined via the rate distributions $\varphi_{j,k}^{\zeta}(t)$ which in turn, as seen in (6), require the total waiting-time distributions, $\psi_{j,k}(t)$, matters are not entirely straightforward. If one stays with the simplest case ($\psi_{j}^{0}=\psi_{j}^{\beta}=\psi_{j}^{\delta} \equiv 0$) a single {\it extra} parameter proves essential: this might, for example, be taken as the dimensionless  ratio
\begin{equation}
\theta_{j}^{+} = \langle t_{+} \rangle_{j}/\tau_{j} \leq 1.
\end{equation}
But the resulting formula, derived from
\begin{equation}
g_{j}^{+}=\frac{u_{j}}{2 \tau_{j}} \left[\langle t_{+}^{2} \rangle_{j} +\langle t_{-}^{2} \rangle_{j} - 2 \langle t_{+} \rangle_{j}/u_{j} \right],
\end{equation}
which in turn follows with the aid of \mbox{Table I}, has a paradoxical feature (for which reason  we do not quote it). Namely, even if the separate mechanicities, $M_{j}^{+}$ and $M_{j}^{-}$, vanish the parameters $g_{j}^{+}$ and $g_{j}^{-}$  do {\it not}, in general, vanish! The reason is that in a pure kinetic scheme with forward and backward rates $u_{j}$ and $w_{j}$ describing  departures from the same state $j$, the resulting waiting-time distributions, $\psi_{j}^{+}(t)$ and $\psi_{j}^{-}(t)$, share a {\it common} exponential factor, namely, $\exp[-(u_{j}+w_{j}) t]= e^{-t/\tau_{j}}$. Thus even if $\psi_{j}^{+}(t)$ and $\psi_{j}^{-}(t)$ are both simple exponentials, the overall process will {\it not} have a simple chemical description unless the $+$ and $-$ decay rates match. More generally, however, if $M_{j}^{+}=M_{j}^{-}=0$ and the (single) additional condition
\begin{equation}
\langle t_{\pm}^{0} \rangle_{j}=\langle t_{\pm} \rangle_{j} \tau_{j}/(\langle t_{+} \rangle_{j}^{2}+\langle t_{-} \rangle_{j}^{2})
\end{equation}
is met then, indeed,  $g_{j}^{+}$ and $g_{j}^{-}$ vanish. (Similar considerations apply, of course, to the behavior on branches; but the extra process available at a primary state, where a branch starts, must not be forgotten.)

Despite these conceptual complications, one can devise instructive examples with simpler dependence on the mechanicities. One useful case when only forward and backward transitions from a  state $j$ occur is described by
\begin{equation}
\psi_{j}^{\pm}(t)=Q_{j}^{\pm} t^{\nu_{\pm} -1} e^{-\nu_{\pm} t/\tau_{j}},
\end{equation}
where, with (53), the relations
\begin{equation}
Q_{j}^{\pm} = (u_{j}, \ w_{j}) \tau_{j}^{1-\nu_{\pm}} \nu_{\pm}^{\nu_{\pm}}/\Gamma(\nu_{\pm})
\end{equation}
ensure that the normalization condition $\int_{0}^{\infty} (\psi_{j}^{+}+\psi_{j}^{-}) dt=1$ is satisfied. The mechanicities are clearly
\begin{equation}
M_{j}^{+}=1-\nu_{+}^{-1} \quad \mbox{and} \quad M_{j}^{-}=1-\nu_{-}^{-1},
\end{equation}
where, of course, $\nu_{+}$ and $\nu_{-}$ could also depend on the state $j$; the nonexponential parameters are then given by
\begin{equation}
g_{j}^{\pm} = - \case{1}{2}(u_{j}, \ w_{j}) \tau_{j}^{2} (u_{j} M_{j}^{+}+w_{j} M_{j}^{-}).
\end{equation}
This result {\it does}  depend only on the effective rates and the associated mechanicities and the $g_{j}^{\pm}$ do vanish when $M_{j}^{\pm} \rightarrow 0$ in accord with the naive expectations.

Similar examples can be devised when branching and death processes occur from a state $j$. For modeling purposes$^{23}$ we believe the associated loss of full generality is likely to be insignificant. But note that, by (61), the $g_{j}^{\pm}$ are always negative for the special cases considered; if, however, $\nu_{+}=\nu_{-}=1$ so that $M_{j}^{+}=M_{j}^{-}=0$ but there are {\it distinct} time constants, $\tau_{j}^{+} \neq  \tau_{j}^{-}$, then $g_{j}^{+}$ and  $g_{j}^{-}$ will have {\it opposite} signs. We stress, nonetheless, that the results presented in \mbox{Section I} apply for quite general  waiting-time distributions: the mechanicity may  be regarded as an auxiliary concept of intuitive and descriptive value.

\section {\hspace{0.5em}Periodic Sequential Model with General  Waiting Times}

To derive the results presented  in \mbox{Section I}, consider first  the general periodic sequential model with waiting-time  distributions as specified  in \mbox{Fig. 1}. This model can be regarded as a one-dimensional  {\it continuous-time random walk} with $N$ internal states ($j=0,1,\cdots,N-1$), a class of walks considered some decades ago by Montroll and coworkers$^{24-27,29}$. The crucial result, demonstrated by Landman, Montroll and Shlesinger in 1977$^{27}$, is  that the probability $P_{j}(l,t)$ of finding the walker  at site $l$ in state $j$ at time $t$ satisfies the {\it generalized master equation} 
\begin{eqnarray}
\frac{d}{d t}P_{j}(l,t)&=&\int_{0}^{t} \left\{ \varphi_{j-1}^{+}(\tau) P_{j-1}(l,t-\tau) + \varphi_{j+1}^{-}(\tau) P_{j+1}(l,t-\tau) \right. \nonumber\\
 & & \left. - \left[ \varphi_{j}^{+}(\tau)+\varphi_{j}^{-}(\tau) \right] P_{j}(l,t-\tau) \right\} d\tau,
\end{eqnarray}
where the relaxation or memory  functions, $\varphi_{j}^{\pm}(t)$,  are related directly to the  waiting-time distribution functions, $\psi_{j}^{\pm}(t)$, precisely as specified  in \mbox{Eq. (6)}. This master equation replaces the simple kinetic rate equations which were the starting points of the previous analyses$^{17-20,28}$.

For our purposes, we may, without loss of generality, assume that the initial condition is $P_{j}(l,0)=P_{j}^{0} \delta_{l,0}$: i.e., the walker starts at the origin $x=l=0$.  Conservation of probability then dictates
\begin{equation}
\sum_{l=-\infty}^{+\infty} \sum_{j=0}^{N-1} P_{j}(l,t)=1  \quad \quad \mbox{(all $t$)}.
\end{equation}
On the other hand, in  any arbitrary state $j_{l}$ in the kinetic picture the  normalization requirement (4)  yields
\begin{equation}
\int_{0}^{\infty}\left[ \psi_{j}^{+}(t)+\psi_{j}^{-}(t)+\psi_{j}^{0}(t) \right] dt = 1,
\end{equation}
or, in terms of Laplace transforms,
\begin{equation}
 \widetilde{\psi}_{j}^{+}(s=0)+\widetilde{\psi}_{j}^{-}(s=0)+\widetilde{\psi}_{j}^{0}(s=0)  =1.
\end{equation}

In order to find the drift velocity $V$ and dispersion $D$ we now  generalize Derrida's method$^{28}$ by  defining  two auxiliary functions for each state $j$, namely,
\begin{equation}
B_{j}(t) \equiv \sum_{l=-\infty}^{+\infty} P_{j}(l,t),\quad \quad \quad  C_{j}(t) \equiv \sum_{l=-\infty}^{+\infty}(j+N l) P_{j}(l,t).
\end{equation}
The  generalized master equation (62)  then yields
\begin{eqnarray}
\frac{d}{d t}B_{j}(t)&=& \int_{0}^{t} \left\{ \varphi_{j-1}^{+}(\tau) B_{j-1}(t-\tau) + \varphi_{j+1}^{-}(\tau) B_{j+1}(t-\tau) \right. \nonumber \\
 & & \left. - \left[ \varphi_{j}^{+}(\tau)+\varphi_{j}^{-}(\tau) \right] B_{j}(t-\tau) \right\} d\tau.
\end{eqnarray}
Similarly, we obtain
\begin{eqnarray}
\frac{d}{d t}C_{j}(t)&=& \int_{0}^{t} \left\{ \varphi_{j-1}^{+}(\tau) C_{j-1}(t-\tau) + \varphi_{j+1}^{-}(\tau) C_{j+1}(t-\tau) - \left[ \varphi_{j}^{+}(\tau)+\varphi_{j}^{-}(\tau) \right] C_{j}(t-\tau)  \right. \nonumber \\
 & & \left.  - \varphi_{j+1}^{-}(\tau) B_{j+1}(t-\tau) +  \varphi_{j-1}^{+}(\tau) B_{j-1}(t-\tau) \right\} d\tau.
\end{eqnarray}

Again following Derrida$^{28}$  we introduce the ansatz
\begin{equation}
B_{j}(t) \rightarrow b_{j}, \quad \quad C_{j}(t)-a_{j} t \rightarrow T_{j},
\end{equation}
which should be valid when $t \rightarrow \infty$. Because of the  periodicity in $j$  we have
\begin{equation}
b_{j+N}=b_{j}, \quad a_{j+N}=a_{j}, \mbox{ and} \quad T_{j+N}=T_{j}.
\end{equation}

After long times a steady state, $dB_{j}/dt=0$, will be achieved.  Then,  recalling in (35) that $\int_{0}^{\infty} \varphi_{j}^{\pm}(t)dt=\widetilde{\varphi}_{j}^{\pm}(s=0)$,  we  introduce the effective transition rates $u_{j}$ and $w_{j}$  defined, in anticipation,  in   (16). Thus (67) yields
\begin{equation}
0=u_{j-1}b_{j-1}+w_{j+1}b_{j+1}-(u_{j}+w_{j})b_{j}.
\end{equation}
Following precisely the arguments given in \mbox{Ref. 20} [see Eqs. (45)-(47)]  we can then  conclude 
\begin{equation}
b_{j}=\frac{r_{j}}{R_{N}}, \quad \quad \mbox{with} \quad \quad  r_{j}=\frac{1}{u_{j}}\left[ 1+\sum_{k=1}^{N-1} \Pi_{j+1}^{j+k} \right],
\end{equation}
where the compact  notation introduced in  (8) and (15) has been used.

To determine the coefficients $a_{j}$ and $T_{j}$ that control  the behavior of $C_{j}(t)$ in (69)  we substitute this  ansatz  into  (68) concluding, for  $t \rightarrow \infty$,
\begin{eqnarray}
a_{j}&=& \int_{0}^{\infty} \left\{ \varphi_{j-1}^{+}(\tau) [a_{j-1}(t-\tau)+T_{j-1}] + \varphi_{j+1}^{-}(\tau)[ a_{j+1}(t-\tau)+T_{j+1}]  \right. \nonumber \\
 & & \left. - \left[ \varphi_{j}^{+}(\tau)+\varphi_{j}^{-}(\tau) \right] [a_{j}(t-\tau)+T_{j}] - \varphi_{j+1}^{-}(\tau) b_{j+1} +  \varphi_{j-1}^{+}(\tau) b_{j-1} \right\} d\tau.
\end{eqnarray}
Now we may introduce the first moments of the relaxation functions via
\begin{equation}
-\int_{0}^{\infty} \tau \varphi_{j}^{\pm}(\tau) d \tau= (d \widetilde{\varphi}_{j}^{\pm}/d s) | _{s=0}=g_{j}^{\pm},
\end{equation}
[see also (22)] which leads to
\begin{eqnarray}
a_{j}&=& t\left[ u_{j-1}a_{j-1}+w_{j+1}a_{j+1}-(u_{j}+w_{j})a_{j} \right] + \left[  g_{j-1}^{+}a_{j-1}+g_{j+1}^{-}a_{j+1}-(g_{j}^{+}+g_{j}^{-})a_{j} \right] \nonumber \\
 & & +  \left[ u_{j-1}T_{j-1}+w_{j+1}T_{j+1}-(u_{j}+w_{j})T_{j} \right] + \left[u_{j-1}b_{j-1}-w_{j+1}b_{j+1} \right].
\end{eqnarray}
The secular term  proportional to $t$   should vanish here,  which condition requires
\begin{equation}
0=u_{j-1}a_{j-1}+w_{j+1}a_{j+1}-(u_{j}+w_{j})a_{j},
\end{equation}
while the coefficients $T_{j}$ then  satisfy
\begin{eqnarray}
a_{j}&=&  \left[  g_{j-1}^{+}a_{j-1}+g_{j+1}^{-}a_{j+1}-(g_{j}^{+}+g_{j}^{-})a_{j} \right] + \left[ u_{j-1}T_{j-1}+w_{j+1}T_{j+1}-(u_{j}+w_{j})T_{j} \right]  \nonumber \\
 & &  + \left[u_{j-1}b_{j-1}-w_{j+1}b_{j+1} \right].
\end{eqnarray}
Comparing (76) with (71) one can conclude that 
\begin{equation}
a_{j}=A b_{j},
\end{equation}
where, using the normalization $\sum_{j=0}^{N-1} b_{j}=1$ following from (66) and (63), the constant $A$ can be calculated by summing   \mbox{Eqs. (77)} on  $j$: the $T_{j}$ cancel identically which leads to
\begin{equation}
A=\sum_{j=0}^{N-1}a_{j}=\sum_{j=0}^{N-1}(u_{j}-w_{j})b_{j}.
\end{equation}
Then, on using the result (72) for  $b_{j}$, we find 
\begin{equation}
A=N[1-\Pi_{1}^{N}]/R_{N}.
\end{equation}

To obtain  the coefficients  $T_{j}$ we introduce, following  \mbox{Ref 20} [see \mbox{Eqs. (54)-(57)}],
\begin{equation}
y_{j} \equiv w_{j+1}T_{j+1}-u_{j}T_{j},
\end{equation}
and rewrite (75) as
\begin{equation}
y_{j}-y_{j-1}=a_{j}- \left[  g_{j-1}^{+}a_{j-1}+g_{j+1}^{-}a_{j+1}-(g_{j}^{+}+g_{j}^{-})a_{j} \right]-u_{j-1}b_{j-1}+w_{j+1}b_{j+1}.
\end{equation}
The solution of this equation, which is  achieved  using the strategy described in \mbox{Ref. 20}, yields
\begin{equation}
y_{j}=u_{j}b_{j}+(A/N) \sum_{i=0}^{N-1}(i+1)b_{j+i+1}+(a_{j}g_{j}^{+}-a_{j+1}g_{j+1}^{-})+c,
\end{equation}
where $c$ is an arbitrary constant which will cancel in the final formula for the dispersion, $D$  (see \mbox{Refs. 20} and 28). The fact that this expression solves (82) can be checked with the help of the relation
\begin{equation}
u_{j}b_{j}-w_{j+1}b_{j+1}=A/N,
\end{equation}
which follows from \mbox{Eqs. (72) and (79)}. Then, iterating (81) and invoking the periodicity (70) yields the relation
\begin{equation}
T_{j}= \left. -\frac{1}{u_{j}}\left[ y_{j}+\sum_{k=1}^{N-1}y_{j+k} \Pi_{j+1}^{j+k}\right] \right/ (1-\Pi_{1}^{N}),
\end{equation}
which, via (83), (78), (79) and (72), represents an explicit result in terms of the effective rates defined in (16).

Now we can calculate the drift velocity, $V$, and the diffusion constant, $D$, using the  long-time  definitions (1) and (2). The mean position of a particle  is given  by
\begin{equation}
\langle x(t) \rangle=\frac{d}{N}\sum_{l=-\infty}^{+\infty} \sum_{j=0}^{N-1} (j+Nl) P_{j}(l,t)=\frac{d}{N}\sum_{j=0}^{N-1} C_{j}(t).
\end{equation} 
With the aid of the generalized master equation (62) the derivative can be taken which, when $t \rightarrow \infty$, leads to
\begin{equation}
\lim _{t \rightarrow \infty}\frac{d}{dt}\langle x(t)\rangle=\frac{d}{N}  \sum_{j=0}^{N-1}(u_{j}-w_{j})b_{j} = \frac{d}{N}A.
\end{equation}
Using the result (80) yields our  final formula  for the drift velocity, namely,
\begin{equation}
V_{\alpha}=d[1-\Pi_{1}^{N}]/R_{N},
\end{equation}
where we recall that $R_{N}$ is defined in (15). This expression corresponds exactly to   Derrida's original result for the simple sequential  kinetic  model. 

A similar approach suffices  to determine the dispersion. We start from 
\begin{equation}
\langle x^{2}(t) \rangle =\frac{d^{2}}{N^{2}}\sum_{l=-\infty}^{ \infty} \sum_{j=0}^{N-1} (j+Nl)^{2} P_{j}(l,t),
\end{equation}
and again appeal to the master equation (62) in the long-time limit. This leads to
\begin{equation}
\lim_{t \rightarrow \infty} \frac{d}{dt} \langle x^{2}(t) \rangle  =   2 \frac{d^{2}}{N^{2}} \left[ \sum_{j=0}^{N-1}(u_{j}-w_{j})(a_{j}t+T_{j})+\case{1}{2}\sum_{j=0}^{N-1}(u_{j}+w_{j})b_{j} + a_{j}(g_{j}^{+}-g_{j}^{-}) \right]. 
\end{equation}
Then,  using  (86), (87),   and the  definition (2), we obtain
\begin{equation}
D =  \frac{d^{2}}{N^{2}}  \left[ \sum_{j=0}^{N-1}(u_{j}-w_{j})T_{j}+\case{1}{2} \sum_{j=0}^{N-1}(u_{j}+w_{j})b_{j}+ \sum_{j=0}^{N-1} a_{j} (g_{j}^{+}-g_{j}^{-})  -A\sum_{j=0}^{N-1}T_{j} \right].
\end{equation}
The coefficients $T_{j}$ can be re-expressed   using (85) and (83) from which the  constant $c$ then cancels$^{20,28}$. Finally,  the definitions (19) and (20) allow  us to write the dispersion in the form presented  in (17)-(21) while the nonexponential parameters first introduced in (22) are confirmed by (74). Note, that the dispersion consists of two terms, $D_{0}$ and $D_{1}$,   the second arising  purely  from the  deviations  of the  waiting-time distribution functions from  the ``chemical,'' Poissonian  forms.

\section {\hspace{0.5em}Periodic  Model with Branches and   Waiting Times}

Now consider the  one-dimensional periodic model with branches and  waiting-time distributions  as presented in \mbox{Fig. 2}. Let $P_{j,k}(l,t)$ be the probability of finding the walker  at site $l$ in state $j$  of the main sequence (labeled $k=0$) or in state $k=1,\cdots,L$,  on the associated   side branch,  at time $t$. Appealing again to Landman {\it et al.}$^{27}$ this probability is governed by the  generalized master equation 
\begin{eqnarray}
\frac{d }{d t}P_{j,0}(l,t)&=&\int_{0}^{t} \left\{ \varphi_{j-1}^{+}(\tau) P_{j-1,0}(l,t-\tau) + \varphi_{j+1}^{-}(\tau) P_{j+1,0}(l,t-\tau) + \varphi_{j,1}^{\gamma}(\tau) P_{j,1}(l,t-\tau) \right. \nonumber\\
 & & \left. - \left[ \varphi_{j}^{+}(\tau)+\varphi_{j}^{-}(\tau) + \varphi_{j,0}^{\beta}(\tau) \right] P_{j,0}(l,t-\tau) \right\} d\tau,
\end{eqnarray}
for $k=0$, by  
\begin{eqnarray}
\frac{d }{d t}P_{j,k}(l,t)&=&\int_{0}^{t} \left\{ \varphi_{j,k-1}^{\beta}(\tau) P_{j,k-1}(l,t-\tau) + \varphi_{j,k+1}^{\gamma}(\tau) P_{j,k+1}(l,t-\tau)  \right. \nonumber\\
 & & \left. - \left[ \varphi_{j,k}^{\beta}(\tau) + \varphi_{j,k}^{\gamma}(\tau) \right] P_{j,k}(l,t-\tau) \right\} d\tau,
\end{eqnarray}
for $1 \leq k < L$,  while for $k=L$ we have 
\begin{equation}
\frac{d }{d t}P_{j,L}(l,t) = \int_{0}^{t} \left[ \varphi_{j,L-1}^{\beta}(\tau) P_{j,L-1}(l,t-\tau) -  \varphi_{j,L}^{\gamma}(\tau) P_{j,L}(l,t-\tau) \right] d\tau.
\end{equation}
The relaxation functions $\varphi_{j}^{\pm}(t)$, $\varphi_{j}^{\beta}(t)$ and $\varphi_{j}^{\gamma}(t)$, are defined, as before, via their Laplace transforms as specified in (6).   At $t=0$ we may  assume that $P_{j,k}(l;0)=P_{j,k}^{0} \delta_{l,0}$ and  normalization requires 
\begin{equation}
\sum_{l=-\infty}^{\infty} \sum_{j=0}^{N-1} \sum_{k=0}^{L} P_{j,k}(l,t)=1 \quad  \quad  \mbox{(all $t$)}.
\end{equation}
Again following the strategy of Derrida's method as described in the previous section,  we introduce the auxiliary  functions
\begin{equation}
B_{j,k}(t) \equiv \sum_{l=-\infty}^{\infty} P_{j,k}(l,t), \quad  \quad C_{j,k}(t) \equiv \sum_{l=-\infty}^{\infty}(j+N l) P_{j,k}(l,t).
\end{equation}
The time evolution of $B_{j,k}(t)$ is then  described   by the set of $L+1$ equations
\begin{eqnarray}
\frac{d }{d t}B_{j,0}(t)&=&\int_{0}^{t} \left\{ \varphi_{j-1}^{+}(\tau) B_{j-1,0}(t-\tau) + \varphi_{j+1}^{-}(\tau) B_{j+1,0}(t-\tau) + \varphi_{j,1}^{\gamma}(\tau) B_{j,1}(l,t-\tau) \right. \nonumber\\
 & & \left. - \left[ \varphi_{j}^{+}(\tau)+\varphi_{j}^{-}(\tau) + \varphi_{j,0}^{\beta}(\tau) \right] B_{j,0}(t-\tau) \right\} d\tau,  \\
\frac{d }{d t}B_{j,1}(t)&=&\int_{0}^{t} \left\{ \varphi_{j,k-1}^{\beta}(\tau) B_{j,0}(t-\tau) + \varphi_{j,k+1}^{\gamma}(\tau) B_{j,2}(t-\tau)  \right. \nonumber\\
 & & \left. - \left[\varphi_{j,k}^{\beta}(\tau) + \varphi_{j,k}^{\gamma}(\tau) \right] B_{j,0}(t-\tau) \right\} d\tau,  \\
& & \vdots  \nonumber \\
\frac{d }{d t}B_{j,L}(t)& = &\int_{0}^{t} \left[ \varphi_{j,L-1}^{\beta}(\tau) B_{j,L-1}(t-\tau) -  \varphi_{j,L}^{\gamma}(\tau) B_{j,L}(t-\tau) \right] d\tau.
\end{eqnarray}
Similarly, the time evolution of the $C_{j,k}(t)$ obeys the equations 
\begin{eqnarray}
\frac{d }{d t}C_{j,0}(t)&=&\int_{0}^{t} \left\{ \varphi_{j-1}^{+}(\tau) C_{j-1,0}(t-\tau) + \varphi_{j+1}^{-}(\tau) C_{j+1,0}(t-\tau) + \varphi_{j,1}^{\gamma}(\tau) C_{j,1}(l,t-\tau) \right. \nonumber\\
 & & \left.  - \left[ \varphi_{j}^{+}(\tau)+\varphi_{j}^{-}(\tau) + \varphi_{j,0}^{\beta}(\tau) \right] C_{j,0}(t-\tau) +\varphi_{j-1}^{+}(\tau) B_{j-1,0}(t-\tau)\right. \nonumber \\
 & & \left.  - \varphi_{j+1}^{-}(\tau) B_{j+1,0}(t-\tau) \right\} d\tau,  \\
\frac{d }{d t}C_{j,1}(t)&=&\int_{0}^{t} \left\{ \varphi_{j,k-1}^{\beta}(\tau) C_{j,0}(t-\tau) + \varphi_{j,k+1}^{\gamma}(\tau) C_{j,2}(t-\tau)  \right. \nonumber\\
 & & \left. - \left[\varphi_{j,k}^{\beta}(\tau) + \varphi_{j,k}^{\gamma}(\tau) \right] C_{j,0}(t-\tau) \right\} d\tau,  \\
& & \vdots  \nonumber \\
\frac{d }{d t}C_{j,L}(t)& = &\int_{0}^{t} \left[ \varphi_{j,L-1}^{\beta}(\tau) C_{j,L-1}(t-\tau) -  \varphi_{j,L}^{\gamma}(\tau) C_{j,L}(t-\tau) \right] d\tau.
\end{eqnarray}

The previous arguments$^{20,28}$ lead to the expectation
\begin{equation}
B_{j,k}(t) \rightarrow b_{j,k}, \quad \quad C_{j,k}(t)-a_{j,k}t  \rightarrow  T_{j,k},
\end{equation}
for $t \rightarrow  \infty$. At large times the equations of motion (97)-(99) then yield the relations
\begin{eqnarray}
w_{j+1}b_{j+1,0}-u_{j}b_{j,0}&=&w_{j} b_{j,0}-u_{j-1} b_{j-1,0}+(\beta_{j,0}b_{j,0}-\gamma_{j,1}b_{j,1}),\\
\beta_{j,0}b_{j,0}-\gamma_{j,1}b_{j,1}&=&\beta_{j,1}b_{j,1}-\gamma_{j,2}b_{j,2}= \cdots=\beta_{j,L-1}b_{j,L-1}-\gamma_{j,L}b_{j,L}=0,
\end{eqnarray}
where we have invoked the definitions (16) and (25) for the rates $u_{j}$, $w_{j}$, $\beta_{j,k}$ and $\gamma_{j,k}$.   Recalling, likewise,  the definitions (22) and (33),  the coefficients $a_{j,k}$ and $T_{j,k}$ must  satisfy, first, 
\begin{eqnarray}
a_{j,0}&=& \left[g_{j-1}^{+}a_{j-1,0}+g_{j+1}^{-}a_{j+1,0}-(g_{j}^{+}+g_{j}^{-})a_{j,0} \right] + \left[u_{j-1} T_{j-1,0} + w_{j+1} T_{j+1,0}-(u_{j}+w_{j}) T_{j,0} \right]\nonumber\\
& & + \left[ u_{j-1}b_{j-1,0}-w_{j+1}b_{j+1,0} \right] + \left[ g_{j,1}^{\gamma}a_{j,1}-g_{j,0}^{\beta}a_{j,0} +\gamma_{j,1} T_{j,1}  -\beta_{j,0}T_{j,0} \right],  \\
a_{j,1}&=& \left[g_{j,0}^{\beta}a_{j,0}+g_{j,2}^{\gamma}a_{j,2}-(g_{j,1}^{\beta}+g_{j,1}^{\gamma})a_{j,1} \right] + \left[\beta_{j,0}T_{j,0} +\gamma_{j,2}T_{j,2}  -(\beta_{j,1}+\gamma_{j,1}) T_{j,1} \right], \\
\vdots \nonumber \\
a_{j,L}&=& \left[g_{j,L-1}^{\beta}a_{j,L-1}-g_{j,L}^{\gamma}a_{j,L}+ \beta_{j,L-1}T_{j,L-1}-\gamma_{j,L} T_{j,L} \right],
\end{eqnarray}
while the vanishing of the secular terms yields, also, 
\begin{eqnarray}
w_{j+1}a_{j+1,0}-u_{j}a_{j,0}&=&w_{j}a_{j,0}-u_{j-1}a_{j-1,0}+(\beta_{j,0}a_{j,0}-\gamma_{j,1}a_{j,1}),\\
\beta_{j,0}a_{j,0}-\gamma_{j,1}a_{j,1}&=&\beta_{j,1}a_{j,1}-\gamma_{j,2}a_{j,2}= \cdots = \beta_{j,L-1}a_{j,L-1}-\gamma_{j,L}a_{j,L}=0.
\end{eqnarray}
The side-branch functions $b_{j,k}$ are  found easily by solving \mbox{Eqs. (105)} recursively which gives
\begin{equation}
b_{j,k}=\Pi_{j}^{\beta,k} b_{j,0} \quad \quad (k=1, \cdots, L),
\end{equation}
where we have  used the product  notation (9). To  obtain  an  expression for $b_{j,0}$ we follow exactly the method used to derive (72) in  \mbox{Section III}  thereby finding
\begin{equation}
b_{j,0}=r_{j}/R_{N}^{\beta},
\end{equation}
where $r_{j}$ is defined in (15) while  $R_{N}^{\beta}$ was introduced in (24).  Comparing  (104) and (105) with  (109) and (110) leads  to
\begin{equation}
a_{j,k}=A b_{j,k}  \quad \mbox{ with} \quad A=N(1-\Pi_{1}^{N})/R_{N}^{\beta}.
\end{equation}

The final derivation of the expressions for the drift velocity and the dispersion now follows along the lines developed  in the  previous section for the models   without branches. The results have been  presented in full in \mbox{Section I.B}. As regards the velocity the branches generate no changes beyond  the replacement of $r_{j}$ by $r_{j}^{\beta}$ and $R_{N}$ by $R_{N}^{\beta}$. However,  an additional term, $D_{2,\beta}$,  appears in the dispersion: see (31)-(33).

\section {\hspace{0.5em}Periodic  Model with Deaths  and   Waiting Times}

Consider, finally,  the periodic sequential model with waiting-time distributions  and the possibility of an irreversible detachment or death  from each state that is  described by a  waiting-time distribution function  $\psi_{j}^{\delta}(t)=\psi_{j \pm N}^{\delta}(t)$: see \mbox{Fig. 1}.  The  generalized master  equation for the probability $P_{j}(l,t)$ now reads$^{27}$
\begin{eqnarray}
\frac{d}{d t}P_{j}(l,t)&=&\int_{0}^{t} \left\{ \varphi_{j-1}^{+}(\tau) P_{j-1}(l,t-\tau) + \varphi_{j+1}^{-}(\tau) P_{j+1}(l,t-\tau) \right. \nonumber\\
 & & \left. - \left[ \varphi_{j}^{+}(\tau)+\varphi_{j}^{-}(\tau)+\varphi_{j}^{\delta}(\tau) \right] P_{j}(l,t-\tau) \right\} d\tau,
\end{eqnarray}
where, as before, the relaxation  functions $\varphi_{j}^{\pm}(t)$ and $\varphi_{j}^{\delta}(t)$ are related to waiting-time distribution functions via  (5) and (6). We may again  assume that the initial condition is 
\begin{equation}
P_{j}(l,0)=P_{j}^{0} \delta_{l,0} \quad \mbox{ with } \quad \sum_{j=0}^{N-1} P_{j}^{0}=1.
\end{equation}
However, as  discussed in \mbox{Ref. 20}, because  the total  probability is no longer  conserved  [so that $\sum_{l=-\infty}^{+\infty}\sum_{j=0}^{N-1} P_{j}(l,t>0) <1$], we  look for long-time solutions of the generalized   master equation (114) that are  of  the form
\begin{equation}
P_{j}(l,t) \approx e^{-\lambda t -\tau_{j}} \widetilde{P}_{j}(l,t),
\end{equation}
where, as before$^{20}$, the decrement  and the periodic state coefficients, $\tau_{j} \equiv \tau_{j \pm N}$, are to be found  from the requirement  that $\widetilde{P}_{j}(l,t)$ satisfies a suitably  ``renormalized,''  {\it probability conserving}   master equation
\begin{equation}
\frac{d }{d t}\widetilde{P}_{j}(l,t)=\widetilde{u}_{j-1}\widetilde{P}_{j-1}(l,t) + \widetilde{w}_{j+1}\widetilde{P}_{j+1}(l,t)-(\widetilde{u}_{j}+\widetilde{w}_{j})\widetilde{P}_{j}(l,t).
\end{equation}
By substituting the ansatz (116) into the full  generalized master equation (114) we obtain,  for large times, the equation
\begin{equation}
\frac{d }{d t}\widetilde{P}_{j}(l,t)=u_{j-1} e^{\tau_{j}-\tau_{j-1}} \widetilde{P}_{j-1} + w_{j+1} e^{\tau_{j}-\tau_{j+1}} \widetilde{P}_{j+1}-(u_{j}+w_{j}+\delta_{j}-\lambda)\widetilde{P}_{j},
\end{equation}
in which the modified rate  definitions (35) {\it et seq.}, which depend explicitly on $\lambda$,  have been used. Matching terms with those  in  (117) generates the identifications 
\begin{equation}
\tilde{u}_{j}=u_{j}\varepsilon_{j+1}/\varepsilon_{j} \quad \mbox{ and } \quad \widetilde{w}_{j}=u_{j}\varepsilon_{j-1}/\varepsilon_{j}\quad \mbox{ with }\quad \varepsilon_{j}=e^{\tau_{j}}.
\end{equation}
It also yields a  condition which the  $\varepsilon_{j}$ must satisfy for consistency, namely,
\begin{equation}
-w_{j}\varepsilon_{j-1} +(u_{j}+w_{j}+\delta_{j})\varepsilon_{j}-u_{j}\varepsilon_{j+1}=\lambda \varepsilon_{j},
\end{equation}
But, recognizing the periodicity in $j$, this is precisely  equivalent to the eigenvalue equation ${\bf M}{\mbox{\boldmath $\varepsilon$}}=\lambda{\mbox{\boldmath $\varepsilon$}}$, where ${\bf M}$ is  the  $N$$\times$$N$ matrix  specified  in (12) and (13). Since the asymptotic decay is required in (116), $\lambda$ must be the smallest eigenvalue which, clearly, should be real and  positive.

To find  expressions for the drift velocity and the dispersion we now require  three auxiliary  functions, namely,
\begin{equation}
B_{j}(t) \equiv \sum_{l=-\infty}^{\infty} P_{j}(l,t),  \quad  C_{j}(t) \equiv \sum_{l=-\infty}^{\infty}(j+N l) P_{j}(l,t), 
\end{equation}
and also
\begin{equation}
E_{j}(t) \equiv \sum_{l=-\infty}^{\infty}(j+N l)^{2} P_{j}(l,t) .
\end{equation}
For large $t$ we  may expect the  asymptotic behavior
\begin{equation}
B_{j}(t) \approx e^{-\lambda t -\tau_{j} } \widetilde{B}_{j}(t), \quad C_{j}(t) \approx e^{-\lambda t -\tau_{j}} \widetilde{C}_{j}(t),\quad \mbox{ and } \quad \quad E_{j}(t) \approx e^{-\lambda t -\tau_{j} } \widetilde{E}_{j}(t),
\end{equation}
with, extending  Derrida's ansatz, 
\begin{equation}
\widetilde{B}_{j}(t) \rightarrow b_{j}, \quad \widetilde{C}_{j}(t) - a_{j}t \rightarrow T_{j}, \quad \mbox{ and }\quad \widetilde{E}_{j}(t)-e_{j}t^{2}-f_{j}t \rightarrow X_{j}.
\end{equation}
The explicit formulas for the coefficients  $b_{j}$, $a_{j}$, $T_{j}$, $e_{j}$, $f_{j}$ and $X_{j}$ can now  be found straightforwardly  by extending  the procedures outlined in \mbox{Sections III} and IV. However, the detailed calculations are fairly tedious and, because of the presence of the functions $E_{j}(t)$, give rise to the higher order nonexponential parameters, $h_{j}^{\pm}$ and $h_{j}^{\delta}$, defined in (47) and (48). 

The mean displacement at time $t$ must now be suitably normalized so as to include only surviving walkers. Thus we have 
\begin{equation}
\langle x(t) \rangle =\left. \frac{d}{N}\sum_{l=-\infty}^{\infty} \sum_{j=0}^{N-1} (j+Nl) P_{j}(l,t) \right/ \sum_{l=-\infty}^{\infty} \sum_{j=0}^{N-1} P_{j}(l,t)= \left. \frac{d}{N}\sum_{j=0}^{N-1}C_{j}(t) \right/ \sum_{j=0}^{N-1}B_{j}(t),
\end{equation}
while  the desired  mean-square displacement is  similarly given  by
\begin{equation}
\langle  x^{2}(t) \rangle =\left. \frac{d^{2}}{N^{2}}\sum_{l=-\infty}^{\infty} \sum_{j=0}^{N-1} (j+Nl)^{2} P_{j}(l,t) \right/ \sum_{l=-\infty}^{\infty} \sum_{j=0}^{N-1} P_{j}(l,t)= \left. \frac{d^{2}}{N^{2}}\sum_{j=0}^{N-1} E_{j}(t) \right/ \sum_{j=0}^{N-1} B_{j}(t).
\end{equation} 
On  taking the derivatives required by (1) and (2) and the  steady-state limit,  these expressions   yield the results  for the  velocity and the dispersion   given in (34) and (39)-(48). Naturally,  when there is no possibility of detachments [$\psi_{j}^{\delta}(t) \equiv0, \ \lambda=0$] one recovers all the results for the periodic sequential models  with waiting times as reported  in the  Introduction and Summary   and in \mbox{Section III}.

This completes the description of our mathematical analysis. In brief summary, we have introduced linear, periodic sequential stochastic models with general waiting-time distributions and have found explicit expressions for the corresponding mean velocities and dispersions that have been reported in \mbox{Section I}. The simplest sequential models have been extended by including finite branches and by allowing for the possibility of death or detachment processes. The deviations from the exponential waiting-time distribution functions that characterize standard kinetic models embodying Poisson processes, do not change the form of the velocity expressions; however, the dispersions entail nonexponential parameters that enter in a more complicated manner. The concept of ``mechanicity,'' introduced in \mbox{Section II},  is useful to quantify and visualize the departures from the usual ``chemical'' kinetic descriptions.

\section*{Acknowledgments}

The interest of Professor Benjamin Widom has been appreciated.  The support of the National Science Foundation (under Grant CHE 99-81772) is gratefully acknowledged.

\begin{table}[!h]

\centering
\caption{{\bf Expressions for the Rate Moments in Terms of Waiting-Time Moments}. See Eqs. (5)-(7) and note that $v_{n}^{\zeta} \equiv 0$ for $n \geq 1$ when $\psi^{\zeta}(t) \propto \psi(t) \propto e^{-ct}$ for any $c$ ($>0$).}
\vspace{5mm}
\begin{eqnarray*}
 v_{0}^{\zeta}&=&\frac{\mu_{0}^{\zeta}}{\mu_{1}},  \hspace{18mm}  v_{1}^{\zeta}=\frac{\mu_{1}^{\zeta}}{\mu_{1}}-\frac{\mu_{0}^{\zeta} \mu_{2}}{2 (\mu_{1})^{2}},  \\
\\
v_{2}^{\zeta}&=&\frac{\mu_{2}^{\zeta}}{\mu_{1}}-\frac{\mu_{1}^{\zeta} \mu_{2}}{(\mu_{1})^{2}}-\frac{\mu_{0}^{\zeta}\ \mu_{3}}{3(\mu_{1})^{2}}+\frac{\mu_{0}^{\zeta} (\mu_{2})^{2}}{2(\mu_{1})^{3}}.  
\end{eqnarray*}

\end{table}

\newpage
\begin{figure}[!h]
\centerline{\psfig{figure=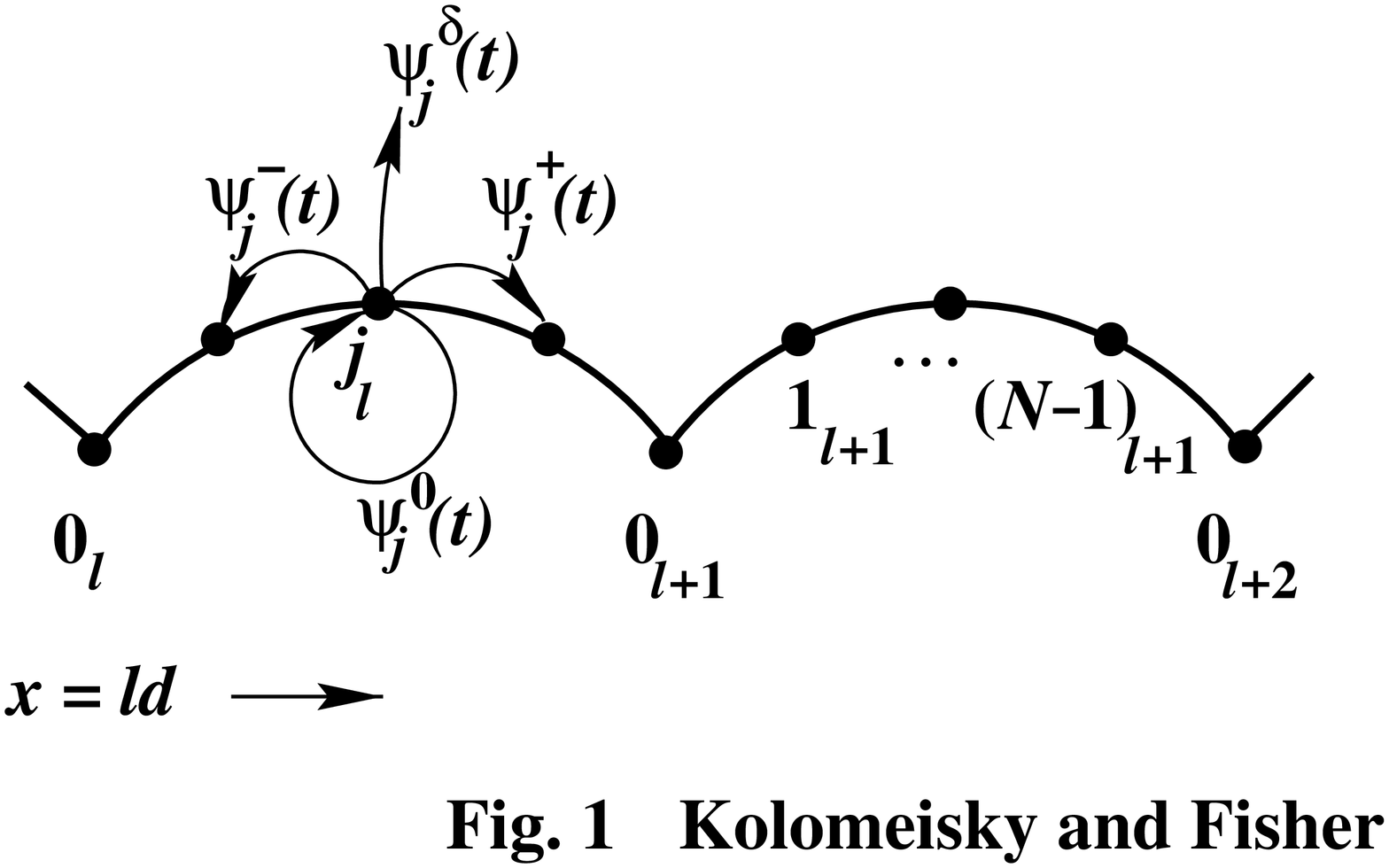,width=13cm,angle=-90}}
\vspace{10mm}
\caption{\hspace{0.4em} A schematic general linear, periodic, stochastic  process with forward and backwards waiting-time distributions, $\psi_{j}^{+}(t)$   and $\psi_{j}^{-}(t)$, failing or futile attempt  distributions $\psi_{j}^{0}(t)$, and irreversible death rates, $\psi_{j}^{\delta}(t)$, from states $j_{l}$, where the reference states, $0_{l}$, are located at positions $x=ld$. (Note that the precise locations of the intermediate states \mbox{$1_{l}, 2_{l}, \cdots, (N-1)_{l}$}, have  no significance for the velocity, $V$, and dispersion, $D$, since these are  defined asymptotically for large times, $t \rightarrow \infty$.) }
\label{fig1}
\end{figure}

\begin{figure} [!h]
\centerline{\psfig{figure=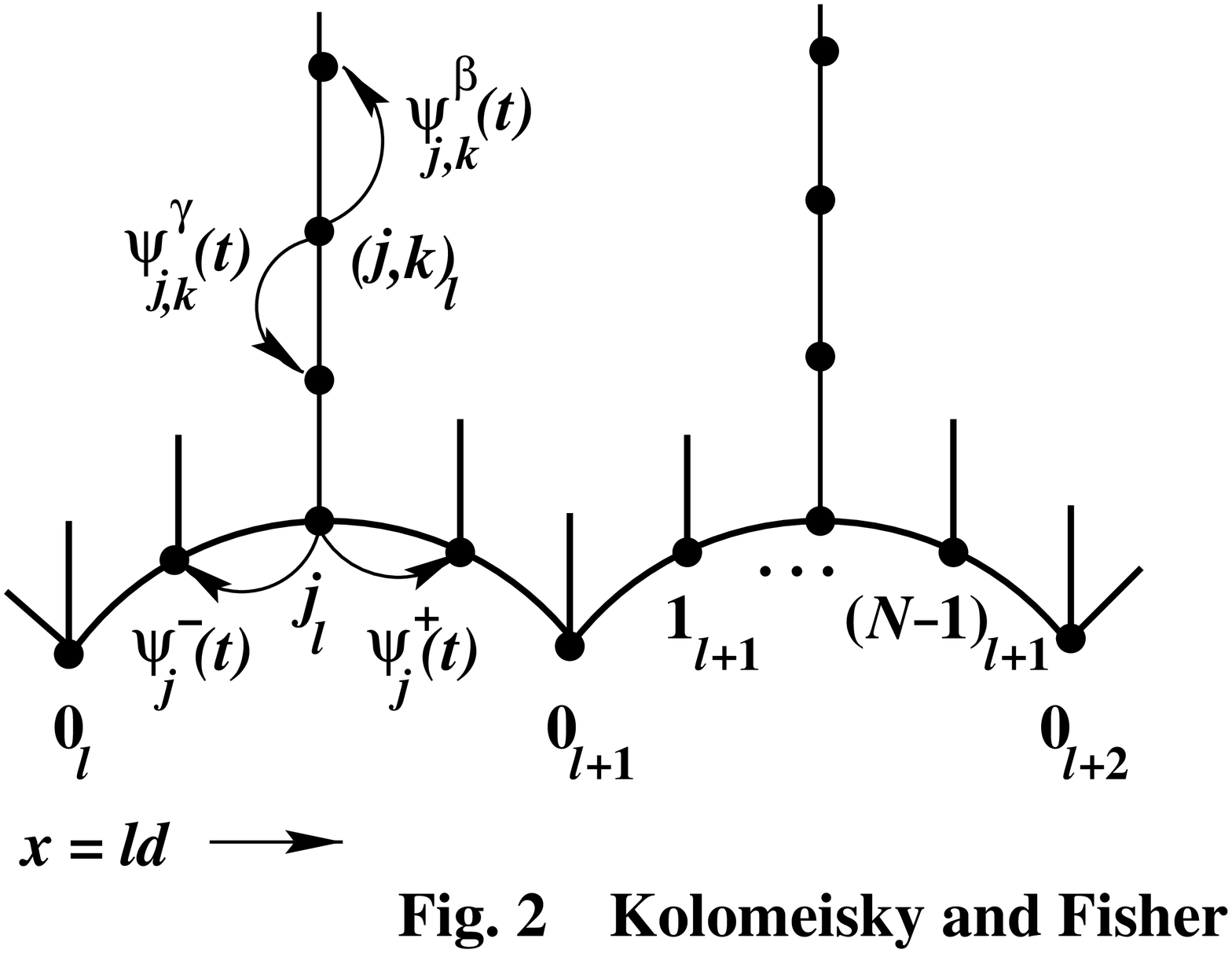,width=11cm,angle=-90}}
\vspace{10mm}
\caption{ \hspace{0.4em} A schematic periodic stochastic  process with branches  of finite length ($\leq L$) grown from each primary site $(j,0)_{l} \equiv j_{l}$;   outward and inward waiting-time distributions, $\psi_{j,k}^{\beta}(t)$ with $k=0,1,\cdots,(L$$-$$1)$, and $\psi_{j,k}^{\gamma}(t)$ with $k=1,2,\cdots,L$, are specified at each   branch site $(j,k)_{l}$. Failing or futile attempt distributions, $\psi_{j,k}^{0}(t)$, are not shown but may be considered as present.}
\label{fig2}
\end{figure}


\begin{thebibliography}{99}
\bibitem{KF}  A.B. Kolomeisky  and M.E. Fisher,  Physica A {\bf 279}, 1 (2000). The following corrections should be noted: in \mbox{Eq. (20)} the coefficient $r_{j}$ should be replaced by $(u_{j}r_{j}/u_{0})$; in  \mbox{Eq. (28)}, which is analogous to  \mbox{Eq. (33)} here, the minus sign   should be deleted; however,  these slips have no affect on the  conclusions or calculations presented.
\end{thebibliography}
\end{document}